\documentclass[aps,prd,prabib,showpacs,nofootinbib,10pt]{revtex4}
\usepackage{graphicx} \usepackage{amsmath} \usepackage{amssymb}
\usepackage{amsfonts} \usepackage{bm}
\usepackage{array}
\usepackage{siunitx}
\usepackage[singlelinecheck=false]{caption}
\usepackage{ytableau}
\usepackage{multirow}
\usepackage{mathtools}

\DeclarePairedDelimiter\floor{\lfloor}{\rfloor}

\begin{document}

\newcommand{\be}{\begin{equation}} \newcommand{\ee}{\end{equation}}
\newcommand{\bea}{\begin{eqnarray}}\newcommand{\eea}{\end{eqnarray}}

\title{Lackadaisical quantum walk  for spatial search}

\author{Pulak Ranjan Giri} \email{pulakgiri@gmail.com}

\affiliation{International Institute of Physics, Universidade Federal do Rio Grande do Norte, Campus Universitario, Lagoa Nova, Natal-RN 59078-970, Brazil} 

\author{Vladimir   Korepin} \email{korepin@gmail.com}

\affiliation{C. N. Yang Institute for Theoretical Physics, State University of New York at  Stony Brook, Stony Brook, NY 11794-3840, USA}

\begin{abstract} 
Lackadaisical quantum walk(LQW) has been an  efficient  technique  in searching  a target state    from a database which is distributed  on a two-dimensional lattice. We  numerically study  the  quantum search algorithm  based on  the  lackadaisical quantum walk on one- and two-dimensions.   It is observed that   specific values of the  self-loop   weight  at each vertex of the graph  is   responsible  for such  speedup of the algorithm. Searching  for a target state on  one-dimensional lattice  with periodic boundary conditions is possible   using  lackadaisical quantum walk, which     can  find  a target state  with    $\mathcal{O}(1)$ success   probability  after   $\mathcal{O} \left( N \right)$ time steps.
In two-dimensions, our numerical simulation   upto $M=6$    suggests  that  lackadaisical quantum walk can search one of the   $M$ target  states    in  $\mathcal{O}\left(\sqrt{\frac{N}{M}\log \frac{N}{M}}\right)$ time steps.
\end{abstract}


\date{\today}

\maketitle 



\section{Introduction} \label{in}
It was suggested by Richard Feynman and  Paul Benioff  in $1982$  that  computations based on the principles  of quantum mechanics  would be more efficient than a  classical computer.  Although  in principle,  a classical computer can simulate a  reasonable size  quantum system,  it   is      not    efficient.  One of  the keys to the success  of building  quantum computers  is  to  have efficient  quantum algorithms  \cite{nielsen,giri} to run on these  devices.  A significant achievement  in this direction  is  realized    when  Peter Shor  in $1994$   \cite{shor1,shor2} showed  that  a quantum algorithm  can  factorize   a large  number  in polynomial-time.   
Then in $1996$  Grover \cite{grover1, grover2,radha1,korepin1,zhang}  came up with an algorithm  which can search for a target element in an unsorted database of size $N$  in    $\mathcal{O}(\sqrt{N})$  time, which is quadratically faster than  the exhaustive classical search of running time $\mathcal{O}(N)$. 
In Grover search,  an initial state is prepared on which    Grover iterator(unitary operator) is acted   repeatedly  until the target state is achieved with significant   success  probability. 
 This can be seen as a rotation of a state in a two-dimensional plane defined by the initial state and the target state.  

Another search problem is  when the elements  of a database  are distributed on  a graph vertices \cite{childs,amba1,amba2,meyer,amba4}. Grover search can be implemented with the help of quantum walks on a complete graph, where each vertex is connected to all other vertices by an  edge and each vertex has a directed self-loop. Grover search can then be seen as a search on both the   spaces of  vertices and edges.
However,   Grover search is not suitable for spatial search.  One such example is a two-dimensional lattice  of  $\sqrt{N} \times \sqrt{N}$ vertices.   The naive argument shows that Grover algorithm will take   
$\mathcal{O}(\sqrt{N})$  iteration  time  to search for a target  state. However each Grover query  needs  $\mathcal{O}(\sqrt{N})$  time to perform  reflection operation. Therefore  the total running time   becomes  $\mathcal{O}(N)$ \cite{beni}, which is  the same as the time  taken by  classical exhaustive search. 
However it has been shown  \cite{am,amba1}   that a recursive  algorithm  combined with amplitude amplification  \cite{brassard} can search a two-dimensional lattice   in    $\mathcal{O}\left(\sqrt{N}\log^2 N\right)$ time, which is better than the exhaustive classical search but less efficient  than the Grover algorithm by a factor of   $\log^2 N$.   For  dimensions  $d \geq 3$  it can search one  target  in    $\mathcal{O} (\sqrt{N})$   time steps  with optimal speed.   

Searching   a graph  by  quantum walks(QW)  \cite{portugal}, which is  the quantum version of classical random walks,  has been an   important technique  in achieving  the   desired   speed.    Since the probability distribution  of quantum walks    spreads quadratically faster than the classical random walks   it is expected  that the  search algorithms based on quantum walks  would outperform  classical search algorithms.  There are two types of quantum walks, namely, discrete and continuous time quantum walks,  both of them can  search a  two-dimensional lattice for a single target in   $\mathcal{O} \left(\sqrt{N} \log N \right)$  time  and   $d \geq 3$-dimensional  lattice  in  $\mathcal{O} \left(\sqrt{N} \right)$  time  \cite{amba2,childs1}.    In the critical case of  $d=2$  dimensions  a  factor  of   
$\mathcal{O} \left(  \sqrt{\log N }\right)$   improvement   in time complexity  has been    achieved     \cite{tulsi, amba3, wong1}.

Recently   lackadaisical quantum walk search   has been introduced  \cite{wong1,wong2,wong3},  which    reduced the   running time of   search on a  two-dimensional  lattice by   a factor of    $\mathcal{O} \left(  \sqrt{\log N }\right)$.    Motivated  by this result  we  in   this article    numerically investigate  lackadaisical  quantum walk search in one and two-dimensional lattice space. 
In particular,  we  will  investigate how much  improvement in  success probability   and  running time    is possible   for    a  search of  a  single target state    in the  one-dimensional  lattice. 
And for two-dimensional lattice,  we   study   the running time and success probability   for  the  search of    multiple target states.  
In the case  of Grover algorithm when the number of target   states   $M$  are increased the running time decreases as  $\mathcal{O}(\sqrt{\frac{N}{M}})$.   Discrete-time quantum walk  search for one of the multiple targets  can be reduced to single target search case, but the running time increases  by a factor of   $\mathcal{O} \left(  \log N \right)$ \cite{amba2}.  For recent works on the   search of  multiple targets  arranged in a cluster  by quantum/lackadaisical quantum walks  see  refs. \cite{rivosh,saha, nahi}.

This article  is  arranged in the following fashion. In section \ref{qw} we briefly discuss  quantum walk search. In section \ref{1D} we study   lackadaisical quantum walk search on one-dimension with periodic boundary conditions.   
In section \ref{2D} we  discuss  lackadaisical  quantum walk search for multiple target states on  a two-dimensional  lattice   and  finally in section \ref{con} we conclude.

\begin{figure}[h!]
  \centering
     \includegraphics[width=1.0\textwidth]{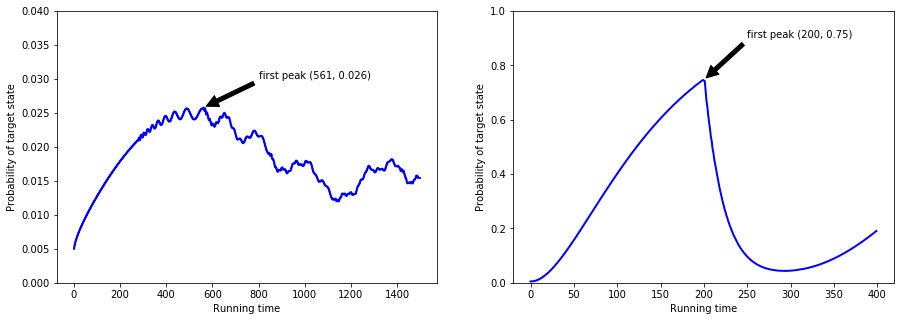}
          
       \caption{(Color online)Variation of  probability of the target state  on a line of $N = 200$ elements  with respect to the running time  $T$.  (a) First  peak of the success  probability   $\approx 0.026$ is  achieved  at   $T= 561$  for a  symmetric coin  with   $\gamma= 0.5$ for the coin of the non-target states and  $\gamma = 0.4$ for the coin of the target state.
(b) First  peak of the success  probability    $\approx 0.75$ is   achieved  at   $T= 200$  for a  lackadaisical  quantum walk of   weight   $a= 2/N $ of the self-loop.}

\end{figure}

\section{ Quantum walk   search on  graph} \label{qw}
Here we briefly discuss how  discrete-time quantum walk can be exploited to search on a graph. We consider a  Cartesian graph $G(V, E)$, however the discussion can be generalized to other graphs as well.   The space of vertices $\mathcal{H}_V$  with dimensions $N$ is the the space of database where we want to define our quantum search.  On $d$-dimensional  Cartesian graph with periodic boundary conditions each vertex is attached to  $2d$-edges. A walker on a vertex can move to $2d$  possible direction by one step at a time. Similar to the coin tossing in classical random walk here we have  a   $2d$-dimensional   coin space  $\mathcal{H}_C$. So the quantum walk is defined on the tensor product space  $\mathcal{H} = \mathcal{H}_C \times \mathcal{H}_V$.   For  the  search algorithm   we usually  start with an initial state which is an equal superposition of the basis states  on both the  coin and vertex space 
\begin{eqnarray}
|\psi_{in}\rangle =  |\psi_c \rangle \otimes  |\psi_v \rangle  =  \frac{1}{\sqrt{2d}} \sum^{d}_{i =1} \left( |x_c^{i+}\rangle + |x_c^{i-}\rangle \right)  \otimes \frac{1}{\sqrt{N}} 
\sum^{\sqrt[d]{N}-1}_{x_v^1, x_v^2, \cdots, x_v^d=0} |x_v^1,  x_v^2, \cdots,  x_v^d  \rangle \,,
\label{in}
\end{eqnarray}
where   $|\psi_c \rangle$ and  $|\psi_v \rangle$ are  the basis states of the coin space and  vertex space respectively.  We need to apply a suitably chosen unitary operator $\mathcal{U}$ on $|\psi_{in}\rangle$  till it  reaches sufficiently close to the target state we are searching for.  For the quantum walk 
\begin{eqnarray}
 \mathcal{U} = S C\,, 
\label{uqw}
\end{eqnarray}
where $C$ is the coin operator which acts on the coin space as an unitary operator.   Hadamard coin and Grover coin  are some  coin operators  used in quantum  walks.   However, for the quantum search  we have to distinguish   the target states   $|t_v \rangle$  from  the non-target states, which we can do by  applying  two  different  coins, one to the target states and other to the non-target states.  Another  useful way is to  apply the same coin  $C_1$   on both types of states but with opposite signs 
\begin{eqnarray}
C=  C_1 \otimes \left( \mathbb{I} -   2 |t_v \rangle \langle t_v | \right)\,, 
\label{qc}
\end{eqnarray}

The shift operator $S$  moves  the walker from one vertex to its immediate  nearest neighbor  vertices connected by $2d$-edges. However this shift operator is not efficient for search. Another one is the flip-flop  shift operator, which   besides shifting the walker from one vertex to the other it also inverts the direction as 
\begin{eqnarray} \nonumber
S =  \sum^{ \sqrt[d]{N}-1}_{x_v^1, x_v^2, \cdots, x_v^d=0} \sum^{d}_{i=1}  |x_c^{i-} \rangle \langle x_c^{i+} | \otimes |x_v^1,  x_v^2, \cdots,  x_v^i+1, \cdots,  x_v^d  \rangle \langle x_v^1,  x_v^2, \cdots,  x_v^d  | \\ 
+  |x_c^{i+} \rangle \langle x_c^{i-} | \otimes |x_v^1,  x_v^2, \cdots,  x_v^{i}-1, \cdots,  x_v^d  \rangle \langle x_v^1,  x_v^2, \cdots,  x_v^d  |\,.
\label{fshift}
\end{eqnarray}
As mentioned before in quantum  search     the final state   is obtained  after repeated application  of   $\mathcal{U}$   to the initial state 
\begin{eqnarray}
|\psi_{f1 }\rangle =    \mathcal{U} ^{T_1} |\psi_{in }\rangle  \,.
\label{fi}
\end{eqnarray}
If    $ |\langle t_v |\psi_{f1 }\rangle|   \approx 1$, we reached  to  the target state with hight  fidelity  with the  time complexity of the algorithm  being  $T_1$.     In cases,  where    $ |\langle t_v |\psi_{f1 }\rangle|   =  1/ T_2  << 1 $,   amplitude  amplification   technique of Grover  is used    $T_2$ times  to amplify the  amplitude of  the  target state  to  $\mathcal{O}(1)$  in  the evolving  state.  So, the total time complexity becomes    $T=  T_1 T_2$. 
For the two-dimensional  lattice   $T_1 = \mathcal{O}(\sqrt{N\log N})$ and  $T_2 = \mathcal{O} \left(\sqrt{\log N}\right)$,  total time complexity  is   $T = \mathcal{O}(\sqrt{N}\log N)$
 with $\mathcal{O}(1)$  success probability   of the target  state.

\section{ LQW  search  in one-dimensional lattice } \label{1D}
A one-dimensional lattice of size   $N$   with periodic boundary conditions is  the simplest model for search by quantum walk.  The standard deviation for  the quantum walk   $\sigma{(t)}= t$  is quadratic  compared to the standard deviation  $\sigma{(t)} = \sqrt{t}$  of the corresponding   classical  random  walk with    after $t$ time steps.   This   ballistic  spread of probabilities  is   a  possible  indication that the quantum walk   would be  faster than the classical walk.     In the worst case  scenario  exhaustive  classical  search algorithm can find  a target  state   in   $\mathcal{O}(N)$   time.    A  classical   algorithm based on  the  random  walk  can find a target  state   in    $\mathcal{O}(N^2)$ time. However a quantum algorithm based on quantum walk   finds a target  state  in  $\mathcal{O}(N)$   time but with  $1/N$  success  probability,  which is not at all any improvement  from the   success probability of the target  state  in  the initial  state.    Numerical  simulation   in ref.  \cite{lovett}  for  $N= 50$   vertices  on a line with periodic boundary conditions shows that  the peak of   success probability of  about  $2\pi/N$  is  obtained for a  search  of  single target  state  after about  $5N$  time.   They used  a symmetric   version of  the Hadamard coin  $H_\gamma^1$  with  $\gamma = 0.45$   for the transformation of     the target   state  and  $\gamma =0.5$ for the transformation of the non-target states.  

Since the  lackadaisical quantum walk improves  the efficiency  of  searching  on  two-dimensional   lattice \cite{wong3}, we   discuss how it  works for a  search on a   one-dimensional  lattice. 
We add one self-loop on each vertex of the lattice  with a specific weight.    The Hilbert space of the quantum coin    $\mathcal{H}_C$   is  spaned by  the basis states    $| x_c^- \rangle,  | x_c^+ \rangle, | x_c^0 \rangle$   and  the Hilbert space of the vertices     $\mathcal{H}_V$  is  given the basis states     $|x_v \rangle$;  $x_v \in [0,   N-1]$.
The quantum walker can go  one step to the left, one step to the right  or  can  stay  in the same position   as
\begin{eqnarray}
S | x_c^- \rangle \otimes | x_v \rangle  &=&    | x_c^+ \rangle \otimes | x_v-1 \rangle\,, \\
S | x_c^+ \rangle \otimes | x_v \rangle   &=&    | x_c^- \rangle \otimes | x_v + 1 \rangle\,,\\
S | x_c^0 \rangle \otimes | x_v \rangle  &=&    | x_c^0 \rangle \otimes | x_v \rangle\,,
\label{flip}
\end{eqnarray}
where  $S$  is the flip-flop shift operator,   which can also  be obtained  from  eq. (\ref{fshift})     by  putting $d=1$.   Quantum search starts with an  initial state  
\begin{eqnarray}
|\psi_{in}^{1D}\rangle =    |\psi_c^{1D} \rangle \otimes \frac{1}{\sqrt{N}} \sum^{N-1}_{x_v=0} |x_v\rangle \,,
\label{instate}
\end{eqnarray}
where  the coin state   $|\psi_c^{1D}\rangle$    is   a weighted superposition of the  coin basis 
\begin{eqnarray}
|\psi_c^{1D}\rangle =    \frac{1}{\sqrt{2+ a}} \left(  | x_c^- \rangle + | x_c^+  \rangle + \sqrt{a}| x_c^0 \rangle \right)\,.
\label{cstate}
\end{eqnarray}
For the lackadaisical quantum walk    we first need   to   rotate the  coin state   $|\psi_c^{1D}\rangle$    by   a coin operator   then   the  vertex state is evolved  by the  flip-flop shift operator  $S$.   We choose the Grover diffusion operator  
\begin{eqnarray}
C_G= 2 |\psi_c^{1D} \rangle \langle \psi_c^{1D} | - \mathbb{I}_3\,, 
\label{cgrov}
\end{eqnarray}
for rotation of the coin state.  To  perform  a search  by  LQW   we have to   devise  a way to recognize  the  target  state from the rest of the states, which we  do by modifying the coin operator.   We  use   $C_G$   for    all basis states of  vertex space except  for the  target state  $| t_v\rangle$ in which case we use  $- C_G$.  We   can    express  this modified  coin operator  as  a single  quantum coin operator  as 
\begin{eqnarray}
\tilde{C_G}=  C_G  \otimes \left(  \mathbb{I}_N  - 2|t_v\rangle \langle t_v|\right)\,. 
\label{mcgrov}
\end{eqnarray}
Note that  the operator    $\tilde{C_G}$  acts  not only on the coin space but also on the vertex space.  The initial state  evolves  under the repeated application of the evolution operator 
\begin{eqnarray}
 \mathcal{U}_{1D} = SC_G \otimes \left(  \mathbb{I}_N  - 2|t_v\rangle \langle t_v|\right)\,. 
\label{mcgrov}
\end{eqnarray}

\begin{figure}[h!]
  \centering
    \includegraphics[width=1.0\textwidth]{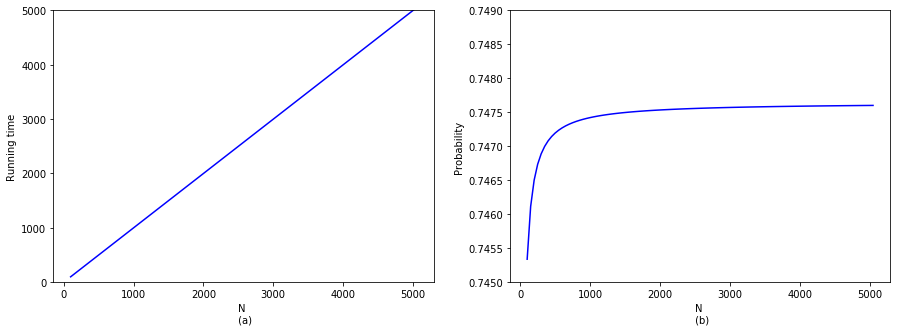}
    
 \caption{(Color online) (a) Running time,  and    (b) corresponding  first  peak of the success probability  for one target  with   $a= 2/N$. }
\end{figure}
However the state thus obtained 
\begin{eqnarray}
|\psi_{1D }\rangle =    \mathcal{U}_{1D} ^{T_1} |\psi_{in}^{1D}\rangle  \,,
\label{1Df}
\end{eqnarray}
has very  low success probability in general for  quantum walk searches with no self-loop. Slightly better success probability can be achieved by suitably  choosing the coin operators as has been shown in ref. \cite{lovett} by using  biased Hadamard coin  or symmetric version of the Hadamard coin

\begin{eqnarray}
H_{\gamma}^\delta = 
\left(\begin{array}{cc} \sqrt{\gamma} & (-1)^{\delta/2}\sqrt{1-\gamma}\\
(-1)^{\delta/2}  \sqrt{1-\gamma} &  (-1)^{1+\delta} \sqrt{\gamma} \end{array}\right)\,.
\label{symh}
\end{eqnarray}
For $\delta =0$  eq. (\ref{symh}) is a  biased Hadamard coin $H_{\gamma}^0$  and for  $\delta =1$(positive square-root)  it is a symmetric Hadamard coin  $H_{\gamma}^1$.  In FIG. 1(a) we have  shown 
the variation of the success  probability of the target state  on a line of size $N = 200$  with respect to the running time  $T$.   First  peak of the success  probability of $\approx 0.026$ is reached at   $T= 561$  for a  coin $C_{sym}$ which acts as a  symmetric Hadamard  coin  $H_{0.5}^1$  on  the non-target states and  $H_{0.4}^1$  on   the target state
\begin{eqnarray}
C_{sym}=   H_{0.5}^1\otimes \left( \mathbb{I} -   |t_v \rangle \langle t_v | \right) +  H_{0.4}^1\otimes |t_v \rangle \langle t_v |\,. 
\label{qcsym}
\end{eqnarray} 

Numerical study on  search  by lackadaisical quantum walk   shows more promising results.  The  first  peak of the success  probability of $\approx 0.75$ is reached at   
$T= 200$  for a self-loop  of   $a= 2/N$  in FIG. 1(b).  The first peak of the success probability of  $ \approx 75$ percent(FIG. 2(b))  is reached in time $N$(FIG. 2(a))   for long range of the sizes   $N \in  \left[100, 5000\right]$  of the  one-dimensional lattice.  Note that success probability  of  $ \approx 75$  percent  is almost constant for large    $N$  and   is  fairly  large compared to the search by  regular  quantum walk  without self-loop.   So the   algorithm  can find  the target state     in  $\mathcal{O}(N)$ time  steps.

\begin{figure}[h!]
  \centering
    \includegraphics[width=1.0\textwidth]{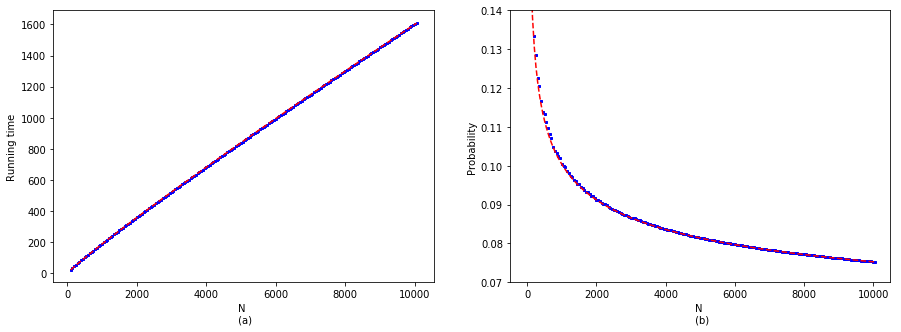}
    
             \caption{(Color online) (a) Running time(blue square)   and  curve fit  by the   red dashed curve    $ 0.30485313 N^{0.93010129}$      with  their corresponding   (b) success probability  $\mathcal{O} \left(1/\log N  \right)$(blue square)  and its  curve fit by red dashed curve   for one target  with the weight of   self-loop   $a= 2/N$  for  $\sqrt{N} \in \left[100,  150, \cdots, 10050\right]$.}
\end{figure}

Alternatively one can think of  applying  lackadaisical quantum walk  till the  success probability first reaches to   $p_s= \mathcal{O}(1/\log N)$  and then  apply      amplitude amplification     \citep{brassard}        $1/\sqrt{p_s}$ times  to  get success probability of    $\mathcal{O}(1)$.   
In FIG. 3(b) blue square  curve  is  the   success probability   when it first crosses the $\mathcal{O}(1/\log N)$  mark  shown against the numbers of  vertices $N$ on the lattice.   The corresponding 
running time is plotted as blue square  curve   in   FIG. 3(a).  From the curve fitting  it  shows that  the running time  is   $\mathcal{O}(N^{0.93010129})$ for  a  self-loop   $a= 2/N$.  
Since  $\mathcal{O}(1/\log N)$ is a low success probability for large lattice size $N$  one has to apply  amplitude amplification  $\mathcal{O} \left( \sqrt{\log N} \right)$ times  to achieve  $\mathcal{O}(1)$ success probability, which increases  the time complexity to    $ T= \mathcal{O}(N^{0.93010129}  \sqrt{\log N} )$.    By  a  suitable  curve  fit  again    we obtain    
 $ T= \mathcal{O}(N^{0.98212135 } )  \simeq  \mathcal{O}(N)$, which  is in good agreement    with the    result  of  the running time    $\mathcal{O}(N)$ obtained  in   FIG. 2  using only lackadaisical quantum walk.

\section{LQW  search  in two-dimensional lattice} \label{2D}
On a   periodic $2$-dimensional   lattice    of size   $\sqrt{N} \times \sqrt{N}$, there are  $N$    vertex    points    $(x, y)$ and each vertex has  $4$-edges. The  collection  of  vertices is  basically the unsorted database  in our case and form the Hilbert space of vertices $\mathcal{H}_V$ with the basis states   $| x_v^1, x_v^2 \rangle$.   We add one self-loop   to  each vertex so the coin space has $5$ dimensions. 
The initial state for the purpose of the  lackadaisical quantum walk is given by 
\begin{eqnarray}
|\psi_{in}^{2D}\rangle =  |\psi_c^{2D} \rangle \otimes  \frac{1}{\sqrt{N}} 
\sum^{\sqrt{N}-1}_{x_v^1, x_v^2=0} |x_v^1,  x_v^2\rangle \,,
\label{2Din}
\end{eqnarray}
where     the coin state     is 
\begin{eqnarray}
|\psi_c^{2D} \rangle   =  \frac{1}{\sqrt{4+ a}} \left( |x_c^{1+}\rangle + |x_c^{1-}\rangle+ 
 |x_c^{2+}\rangle + |x_c^{2-}\rangle  + \sqrt{a}| x_c^0\rangle \right) \,.
\label{2Dcstate}
\end{eqnarray}

\begin{figure}[h!]
  \centering
    \includegraphics[width=1.0\textwidth]{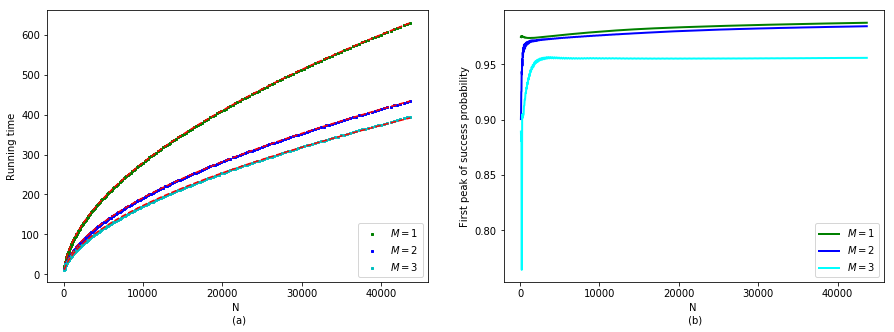}
    
             \caption{(Color online) Running time(a)  with  their corresponding  first  peak of the success probability(b)  for one target(green) with   $a= 4.01/N$ , two targets(blue)  with   $a= 7.8/N$       and three targets(cyan)  with   $a= 10.4/N$    respectively for $\sqrt{N}= 10, 11, \cdots, 209$.}  
\end{figure}

For the lackadaisical quantum walk    we first need   to   rotate this   coin state      by   a coin operator, which is   followed by    flip-flop shift operator  to  evolve  the vertex state.   We choose the Grover diffusion operator  
\begin{eqnarray}
C_{2D}= 2 |\psi_c^{2D} \rangle \langle \psi_c^{2D} | - \mathbb{I}_5\,, 
\label{2Dcgrov}
\end{eqnarray}
for  the rotation of the coin state.  To   recognize  the  target  state from the rest of the states   we  modify  the coin operator 
\begin{eqnarray}
\tilde{C_{2D}}=  C_{2D} \left(  \mathbb{I}_N  - 2|T_M\rangle \langle T_M|\right)\,, 
\label{2Dmcgrov}
\end{eqnarray}
where  $|T_M \rangle =  1/M \left(  |t_1\rangle   +  |t_2\rangle + \cdots,  |t_M\rangle   \right)$  is the equal superposition of  $M$ target states.   
The  initial state    after repeated application of    $ \mathcal{U}_{2D} $   becomes 

\begin{eqnarray}
|\psi_{2Df }\rangle =    \mathcal{U}_{2D} ^{T_1} |\psi_{2Din}\rangle  \,,
\label{2Df}
\end{eqnarray}
where     $ \mathcal{U}_{2D} =  S_{d=2}\tilde{C_{2D}}$.  The shift operator  $S_{d=2}$  can be readily obtained  by putting $d=2$ in eq.  (\ref{fshift}). 

\begin{figure}[h!]
  \centering
   \includegraphics[width=1.0\textwidth]{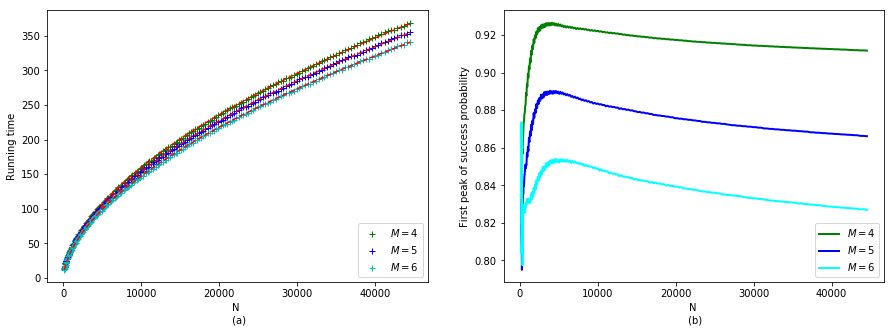}
        
             \caption{(Color online) Running time(a)  with  their corresponding  first  peak of the success probability(b)  for four  target(green) with   $a= 15.2/N$ , five targets(blue)  with   $a= 18.6/N$       and six  targets(cyan)  with   $a= 21.7/N$    respectively for $\sqrt{N}= 12, 13, \cdots, 211$.}  
\end{figure}

In FIG. 4(a)  numerical simulations     for the running time   for the cases   $M=1$(green square  curve),    $M=2$(blue square  curve)  and    $M=3$(cyan  square  curve)   has been shown for    the   two-dimensional lattice  size   $\sqrt{N} = \left( 10,  11, \cdots,  209 \right)$  
and their corresponding   success probability to find one of the  target states   has been plotted in  FIG. 4(b).   
Best  fit   for the running time shown by the red dashed curves in FIG. 4(a)     for $M=1, 2, 3$  targets  are   given by
\begin{eqnarray}\nonumber 
T_1&=& 0.76766755\sqrt{N\log N}\,, \\ \nonumber
T_2&=& 0.773523 \sqrt{\frac{N}{2}\log \frac{N}{2}}\,, \\ 
T_3&=& 0.87265627 \sqrt{\frac{N}{3}\log \frac{N}{3}} \,,
\label{bestfit1}
\end{eqnarray}
respectively  for   three target states   $|\floor{\sqrt{N}/2},   \floor{\sqrt{N}/2} \rangle,  | 2, 2 \rangle$  and   $|7, 7 \rangle  $.     We have also   studied   running time  and  sucess probability  for $M=4, 5$  and $6$ targets  in FIG. 5   where   $|\floor{\sqrt{N}/2},   \floor{\sqrt{N}/2} \rangle,  | 2, 2 \rangle$,    $|7, 7 \rangle$,
$|4, 4 \rangle$,  $|8, 8 \rangle$ and  $|10, 10 \rangle$  are the target states. 
Best  fit   for the running time shown in  the red dashed curves in FIG. 5(a)     are   given by
\begin{eqnarray}\nonumber 
T_4&=& 0.95206188\sqrt{\frac{N}{4}\log \frac{N}{4}}\,, \\ \nonumber
T_5&=& 1.03816497\sqrt{\frac{N}{5}\log \frac{N}{5}}\,, \\ 
T_6&=& 1.10334645 \sqrt{\frac{N}{6}\log \frac{N}{6}} \,,
\label{bestfit2}
\end{eqnarray}
The pre-factor for the running time  in eqs.  (\ref{bestfit1})   and   (\ref{bestfit2})   increases  slightly   when the number of targets increase.    Our numerical simulation  results  for multi target spacial search   upto $M=6$   presented in  eqs. (\ref{bestfit1})  and (\ref{bestfit2})  suggests  that  lackadaisical quantum walk can search one of the   $M$ target  states    in  
\begin{eqnarray}
T = \mathcal{O}\left(\sqrt{\frac{N}{M}\log \frac{N}{M}}\right)\,,
\label{bestfit}
\end{eqnarray}
 time steps.  Note that  we have chosen   different  values for the  self-loop weight  $a$  for  different number of target states keeping  large lattice size in mind, since  we are interested  in time complexity for large $N$.   
In the standard analysis of lackadaisical quantum walk search,  the value   of   $a$  for which the  first peak of the success probability  is maximum, is  chosen   for the   evaluation of the running time  as   has been   presented    in ref.   \cite{wong3}.  This critical  value  of  $a$   usually  depends on the number of  targets,  the size of the lattice,   the  particular  graph  and its degree on which the   quantum search is  carried out.      
The  variation of the success  probability   to find  one of the  target states    as a function of  $Na$  is  shown  in  FIG. 6  for  fixed   $N$ and $M$.      
\begin{figure}[h!]
  \centering
    \includegraphics[width=1.0\textwidth]{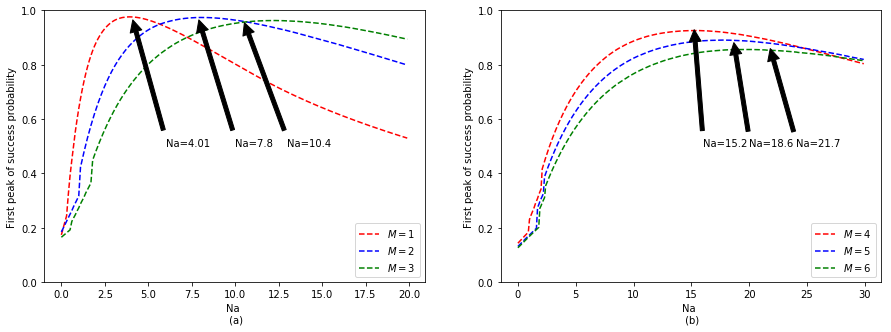}
    
             \caption{(Color online)  (a) Variation of the first peak of the success probability  for $M=1$(red), $M=2$(blue) and $M=3$(green)  targets and (b) variation of the first peak of the success probability  for $M=4$(red), $M=5$(blue) and $M=6$(green)  targets  as a function of  $Na$  for  $70 \times 70$ square lattice . }  
\end{figure}
The  success probabilities are  close to the peaks  for the values   of  $a$(self-loop)   we have chosen  in our analysis as can be seen  from  FIG. 6.   
Note that  the success probabilities   bellow  about   $N= 2500$  slightly   drop   in FIG. 4 and 5, because  our choices of the  self-loop weight are not optimal   for small  lattice size.

\section{Conclusions} \label{con}
Quantum walk  has been an important tool for developing  quantum search  algorithms.   On two-dimensional   lattice,  it has been observed  that the  regular quantum walk(no self-loop) searches  a target state  with  
$\mathcal{O} \left( \frac{1}{\log N}\right)$    success probability   in    $\mathcal{O} \left(\sqrt{N \log N}\right)$  time steps.   Since this success probability is not significant  we need to  exploit  amplitude amplification technique of  Grover to get  constant   $\mathcal{O}(1)$ success probability  at the cost of   an increase of the  running  time   by a factor of  $\mathcal{O} \left(\sqrt{\log N} \right) $.      The total time complexity  is therefore  $\mathcal{O} \left(\sqrt{N}\log N \right)$.   One can  improve  the  running time  by  introducing  an ancilla   qubit  and modifying the quantum walk search  \cite{tulsi}  so that    constant   $\mathcal{O}(1)$   success probability   is achieved  in   $\mathcal{O} \left(\sqrt{N \log N}\right)$  time steps  without  involving    amplitude amplification.  Another approach is to consider a small neighborhood of the target state \cite{amba3}, whose total  probability is  $\mathcal{O}(1)$ without further performing     amplitude amplification. 

Improvement of the running time  has also  been  attributed to the laziness of the quantum walk in a recent work  \cite{wong3} where a self-loop is attached on each vertex of the two-dimensional lattice. This approach, known as the  lackadaisical quantum walk, has generated some interests  in the critical case of $d=2$ dimensional lattice.  We have studied the effect of laziness to   search  a target on a  one-dimensional periodic lattice.  It is known that  quantum walk search in one-dimensional lattice is inefficient. However,  our study shows that using the lackadaisical quantum walk we can  increase the efficiency of the algorithm.  In $N$ time steps it is possible to achieve a constant        $\mathcal{O}(0.75)$  success probability   exploiting only lackadaisical quantum walk with  self-loop  weight $a=2/N$(laziness), which is much better than to have a peak success  probability of 
$2\pi/N$  after $2N$  or more time steps \cite{lovett} by  using  regular  quantum walk for search.  
Lackadaisical quantum walks  followed by amplitude amplification   can find a target  state   with  $\mathcal{O}(1)$   success probability    in    $\mathcal{O}(N)$  time steps.

In two-dimensional lattice we have  studied the effect of lackadaisical quantum walk to search one of the $M$ target states.  The success probability and running time greatly varies  as a function of the  self-loop  weight  $a$   \cite{saha, nahi}.  However for  some suitable choices  of  the self-loop  weight  one of the    $M=1, 2, 3$ target states can be searched in  $\mathcal{O}\left(\sqrt{\frac{N}{M}\log \frac{N}{M}}\right)$ time steps with   $\mathcal{O}(1)$  success  probability.  We found that
$T_1= 0.76766755 \sqrt{N\log N} $,  $T_2= 0.773523 \sqrt{\frac{N}{2}\log \frac{N}{2}} $  and  $T_3= 0.87265627\sqrt{\frac{N}{3}\log \frac{N}{3}} $  are      the best fit with the numerical data  for the  three target states   $|  \floor{\sqrt{N}/2},  \floor{\sqrt{N}/2} \rangle,  | 2, 2 \rangle$  and   $|7, 7 \rangle$.   We have also extended our analysis  up to $M=6$  in FIG. 5.
The choice of these  target states is  random in our numerical simulation, except  for  the cases of so-called  exceptional configurations  \cite{nahi}  of the target states for which there is no speedup.      Note also that  searching  of   $M$  targets arranged   in a   $\sqrt{M} \times \sqrt{M}$   group or  uniformly distributed  with  a spacing of  $\sqrt{N/M}$  between two targets      have  been  studied  \cite{rivosh}, which  shows  that   for the   distributed  case    after   $\mathcal{O}\left(\sqrt{\frac{N}{M}\log \frac{N}{M}}\right)$   time steps    success probability  of   $\mathcal{O} \left(  1/ \log (N/M)\right)$  is achieved   in   regular quantum  walk search.   And for the  grouped  case    $\Omega \left(  \sqrt{N}  - \sqrt{M} \right)$  time steps are required.  The same  system   has   also  been   studied   \cite{saha}    by exploiting  lackadaisical quantum walk, which  gives   better success probability.   However, in our case,   the arrangement  of the targets does not have  any  specific restrictions on  their arrangement.

It would be interesting to extend  our   study to  other graph structures in two-dimensions  to understand how the connectivity   affects    the success probability  and running  time in the presence  of  self-loop.   Lattice  without periodic boundary conditions is also another potential  system to study  search algorithm by lackadaisical quantum walk.

\section*{Acknowledgements} 
P. R. Giri is supported by  International Institute of Physics, UFRN, Natal, Brazil.  V.   Korepin is  grateful to SUNY Center of Quantum Information Science at Long Island for support.


\end{document}